\documentclass{ws-ijqi}
\usepackage{color}
\usepackage{cite}

{\catcode`\|=\active 
  \gdef\Braket#1{\begingroup
\mathcode`\|32768\let|\BraVert\left<{#1}\right>\endgroup}}
\def\BraVert{\egroup\,\mid\,\bgroup}

\newcommand{\e}{\mathcal{E}}

\newcommand{\tr}{\mbox{tr}}

\newcommand{\pd}{\mathfrak{p}}
\newcommand{\qd}{\mathfrak{q}}

\newcommand{\ud}{\mathfrak{u}}
\newcommand{\td}{\mathfrak{t}}

\newcommand{\pt}{\mathbf{P}}
\newcommand{\qt}{\mathbf{Q}}

\begin{document}

\markboth{Vinjanampathy \& Modi}
{Correlations, Operations and the Second Law}

\catchline{}{}{}{}{}

\title{Correlations, operations and the second law of thermodynamics}

\author{Sai Vinjanampathy}
\address{Centre for Quantum Technologies, National University of Singapore\\ 3 Science Drive 2, Singapore 117543.\\
sai@quantumlah.org}

\author{Kavan Modi}
\address{School of Physics, Monash University\\ Victoria 3800, Australia.\\ kavan.modi@monash.edu}

\maketitle

\begin{history}
\received{\today}
\end{history}

\begin{abstract}
Completely positive trace preserving maps are essential for the formulation of the second law of thermodynamics. The dynamics of quantum systems, correlated with their environments, are in general not described by such maps. We explore how this issue can be fixed by describing the classical analogue of this problem. We consider correlated probability distributions, whose subsequent system dynamics is ill-described by stochastic maps, and prescribe the correct way to describe the dynamics. We use this prescription to discuss the classical version of the second law, valid for correlated probability distributions.
\end{abstract}


\section{Introduction}

The laws of thermodynamics are at the cornerstones of our understanding of the physical world \cite{chandler1987introduction}. The zeroth law informs us about the transitivity of equilibria. The first law tells us about the partitioning of internal energy into work and heat. The third law informs us that entropy of a zero temperature system can at best limit to a constant. The second law is a little different from the other three. It is an inequality law that tells us that certain processes, that decrease the entropy of the universe, are not allowed (or unlikely). Stated differently, the second law states that the entropy production of a system that undergoes a cyclical transformation is positive semi-definite, wherein the rest of the universe is returned to its initial state. This law has been at the center of many important theoretical and practical discussions, from designing heat engines to the physics of black holes.

Thermodynamics assumes that the system under study is composed of a large number of constituent subsystems. The presence of such large number of constituent subsystems $N$ means that the fluctuations in certain relevant measurable quantities is reasonably bounded by $N^{-\frac{1}{2}}$, making averages meaningful. Under such conditions, the lack of information about which of all allowed configurations $\Omega$ the system exists in is quantified by the thermodynamic entropy $\mathbb{S}=-k_{B}\log{(\Omega)}$. The second law, in the Clausius form, implies the nonexistence of cyclical transformations such that $d\mathbb{S} < 0$ for the system. Equivalently, it states that for all such cyclical transformations, $d\mathbb{S} \geq 0$. Since such laws are only applicable in the limit of large number of particles, it is of interest to know what generalizations of the second law apply to systems of finite size \cite{lloyd1989use,lloyd2006quantum}. Such generalized laws \cite{ford2006quantum} would be applicable to the arbitrary evolutions of quantum systems, with the additional constraint that they would become the relevant laws of thermodynamics in the limit of large numbers.

Since thermodynamically large systems have proportionally small fluctuations of some observables, a law such as the second law holds true for relevant observables. For small systems relevant to this work and others, the second law still holds on average. This fact however, does not imply that the fluctuations about the average are irrelevant \cite{jarzynski2013equalities}. This is the key observation behind studying the fluctuations of out-of-equilibrium systems. Such fluctuations even for small systems were shown to have the signatures of the deep physical laws which were typically observed for statistically large sizes. For a non-equilibrium process, the fluctuation relation is typically stated as $p_{F}(x)=p_{B}(-x)\exp(\gamma x)$. Here, $p_{F}(p_{B})$ refers to the forward (backward) probability density of an event $x$. The fluctuation relation typically states that events that decrease $x$, and go against the direction set by the second law for equilibrium processes, are generally exponentially suppressed. Employing Jensen's inequality produces an entropy law that can be associated with the second law of thermodynamics. Hence an approach to understand entropy production in quantum systems has been to investigate fluctuation theorems relevant to the system dynamics and conclude entropy production bounds from this \cite{campisi2011colloquium}. Recently the fluctuations relations formalism has made great advances in understanding the thermodynamics out-of-equilibrium biological and chemical processes \cite{west, gupta, england}.

The derivation of the second law from fluctuation relations relies of several assumptions about the dynamics. The system is is assumed to be in microreversible, meaning that the transition probability $A \to B$ in the forward process is the same as the transition probability $B \to A$ in the backward process. Moreover, it is implicitly assumed that the system of interest is not correlated with an hidden system. In this paper we derive a version of the second law for classical stochastic processes that are neither microreversible nor assumed to be independent of any other hidden system. In the next section we discuss the classical limit of the second law for correlated systems. We begin by discussing classical stochastic processes and intermediate correlations of the probability vector of the state with its environment. We then proceed to discuss how to describe the correlated dynamics using preparations and present the second law as a consequence of that analysis. We begin with a brief review of this limit of the second law, which has been generalized in the quantum regime in various ways that are reviewed briefly but non-exhaustively in the rest of this section. 

\section{Second law for quantum processes}

Consider the recent series of works that have explored the relationship between thermodynamics at the mesoscopic scale and quantum correlations \cite{alicki1979quantum, park2013heat, geva1992quantum, abreu2011extracting, kosloff2010optimal, allahverdyan2005minimal, allahverdyan2004maximal, geva1992classical, he2002quantum, erez2008thermodynamic}. One topic of interest has been the design of quantum analogues of engines, refrigerators and other thermal devices \cite{abah2012single, rossnagel2014nanoscale, latifah2011multiple, dillenschneider2009energetics}. Such quantum engines have received considerable attention. Quantum refrigeration has also been investigated, including concrete designs of mesoscopic refrigerators to connections computing. The maximum extractable work from a given quantum state was discussed \cite{allahverdyan2004maximal}, where the authors showed that the non-commuting nature of the density matrix with respect to the Hamiltonian allowed one to extract work from the state, by the means of unitary transformations. Furthermore, extensions of this idea to discuss work extraction via feedback and optimal performance of such engines have also been discussed. Coherence has also been explored in the context of Onsager's relation and a detailed microscopic principle was derived recently \cite{rodriguez2013thermodynamics}. Finally, many works have explored the role of entanglement in extracting work efficiently, see \cite{hovhannisyan2013role} for instance. All of these systems are made of small number of particles and prone to fluctuations. Then how can we construct a second law that applies to thermodynamics at the mesoscopic level? A fluctuation relations approach is not suitable since these systems are often initiated in non-thermal states and experience non-unital (non-microreversible) dynamics.

To understand thermodynamics for system with a small number of degrees of freedom, a description of generic processes is needed. Such a description of generic quantum evolution when the states of the system and environment are uncorrelated is given by completely positive trace preserving (CPTP) maps \cite{nielsen2010quantum, bengtsson2006geometry}. Such maps are defined by their action transforming density matrices to density matrices, and are denoted henceforth by $\Phi$. Such maps are known to admit an operator sum representation, namely $\Phi[\rho]:=\sum_{r} K_{r}\rho \, K^{\dagger}_r$, where $\sum_{r} K^{\dagger}_{r} K_r=\mathbb{I}$. Such maps describe all state dynamics that can be envisioned as arising from a system uncorrelated with an environment interacting with the environment for a time, following by the environment being discarded (traced over). Having outlined the description of generic dynamics, we will discuss how CPTP maps have been used to construct a second law.

One approach to formulate a generalization of the second law for CPTP evolution relates to an important property of quantum relative entropies:
\begin{align}
S(\rho \Vert \sigma)\geq S(\Phi[\rho] \Vert \Phi[\sigma]),
\end{align}
where relative entropy is $S(\rho\Vert\sigma) := \tr\left[ \rho\log(\rho) -\rho \log(\sigma) \right]$. Since we wish to measure the change in entropy of a given state, we choose $\sigma=\mathfrak{e}$, the non-equilibrium steady state that is the fixed point of the CPTP map $\Phi$, namely $\Phi[\mathfrak{e}]=\mathfrak{e}$. This contractivity condition can be re-expressed with this choice of 
\begin{align}
S(\Phi[\rho])-S(\rho)\geq-\mathrm{tr}\left[\Phi[\rho] \log(\mathfrak{e}) -\rho\log(\mathfrak{e})\right],
\end{align}
which expresses a generalization of the second law \cite{spohn1978entropy, alicki1979quantum}. The above expression states how for a given CPTP map $\Phi$, and a given state $\rho$, no transformation that violates the above generalized second law is allowed. Spohn posited this as the generalization of the second law of thermodynamics for semi-group evolution \cite{spohn1978entropy}. Further generalizations of this result involve employing contractivity to R\'{e}nyi divergences have also been considered. This approach to a quantum formulation of thermodynamics involves generalizing von-Neumann entropies, which describes the asymptotic lack of information about the many states composing a quantum mechanical ensemble represented as a state $\rho$, assuming that there are an infinitely many identical copies  of the given state accessible to the experimenter. This is not always the case. In the absence of such infinitely many identical copy conditions, various generalizations of the von-Neumann entropy become important in describing data compression and transmission through a quantum channel, and hence the information-theoretic formulation of quantum thermodynamics. Such a formulation relies on the so-called R\'{e}nyi divergence and the associated R\'{e}nyi relative entropy \cite{brandao2013second, petz2003monotonicity, wilde2013strong}. Such generalizations of the relative entropy has been used to formulate an approach to generalizing the second law of thermodynamics, written in terms of an inequality involving a generalized free energy \cite{brandao2013second}.

To make contact with classical stochastic processes, we note that the issue of entropy production is related to the issue of reversibility. Reversibility of both classical and quantum dynamics has a long standing relationship with conditional states and probabilities. This way, the issue of classical entropy production is one way to investigate the issues of reversibility of dynamics. When the probability vector  describing the state of a system is correlated with an environment that orchestrates stochastic dynamics of the probability vector, the issue of reversibility becomes even more subtle. Whilst earlier attempts at resolving this issue have revolved around a chain of conditional probabilities, in this paper, we resolve this issue by the application of novel tools from quantum information theory. This way, contractivity of relative entropy is shown to have an effect on classical stochastic dynamics, and the effect is to resolve the issue of reversibility. This is detailed in the following sections. We refer the reader to  \cite{vinjanampathy2014second} for the quantum version of the solution presented here, which follows much of the spirit of this derivation.

\section{Classical stochastic processes and intermediate correlations}

In this section we describe a class of problems whose dynamics are not described by maps acting on states due to the presence of intermediate correlations. For this article we will only deal with classical systems subjected to stochastic processes. The quantum case is treated in detail in \cite{modi2012operational, vinjanampathy2014second, arXiv:1410.5826}. The state of a classical system is described by probability distributions and the process is governed by a stochastic matrix. Our aim is to describe scenarios where the  description of the system dynamics in terms of stochastic maps fail. We will point out the specific reasons for this failure and then give a solution yields the full description for the process. The full description involves a mapping from preparations to states. We begin with a brief overview of stochastic processes and the notation in the next subsection.

From chemical reactions \cite{gillespie1977exact} to Brownian motion \cite{van1992stochastic,gardiner2004quantum}, stochastic dynamics has become an integral part of the study of physical systems. Dynamics such as these are treated by analyzing the random variable of interest in terms of their probability distributions. In general, the state of a $d$--dimensional classical system is denoted by a probability distribution. Let us denote the state of the system by the probability distribution $\pd$. For instance, $\pd=(p_{0},p_{1})$ may represent the probability of a two-color coin to either be observed to be red ($p_0$) or yellow ($p_1$). 

To describe how the the probability distributions change in time, a map that describes transitions is needed. For instance, imagine that a coin whose state is $\pd$ enter a machine such that the state of the system changes to $\qd$ upon exiting the machine. Such a transformation is described via a map $\Lambda$ acting on $\pd$ and it is found by inverting the set of linear equations that takes $\pd$ to $\qd$:
\begin{gather}\label{Cstoch}
\Lambda[\pd] = \qd.
\end{gather}
We can think of $\Lambda$ as $d \times d$ matrix that acts on a column vector $\pd$ to yield another column vector $\qd$, both with $d$ entries. In other words, the action of $\Lambda$ on $\pd$ is just matrix multiplication. Suppose we insert a complete set of basis states $\ud_{j}$ into the process and measure the corresponding outputs $\qd_{j}$, then the map is simply
\begin{gather}\label{stochmap}
\Lambda = \sum_j \qd_{j} \ud_{j}^T,
\end{gather}
where the superscript $T$ denotes the transpose of the vector.

We emphasize that $\Lambda$ is independent of the state $\pd$ and describes a property of the machine that transforms $\pd$ to $\qd$. The stochastic map $\Lambda$ represents the actions of the machine with regards to either choosing to let the color remain unaltered as it travels through the machine ($\Lambda_{0;0}$ and $\Lambda_{1;1}$) or changing the color ($\Lambda_{1;0}$ and $\Lambda_{0;1}$). The transition of $p_0$ to $q_0$ is achieved via $\Lambda_{0;0} p_0 + \Lambda_{0;1} p_1$. 

\subsection{Process with intermediate correlations}

\begin{figure}
\begin{center}
\includegraphics[width=0.89\textwidth]{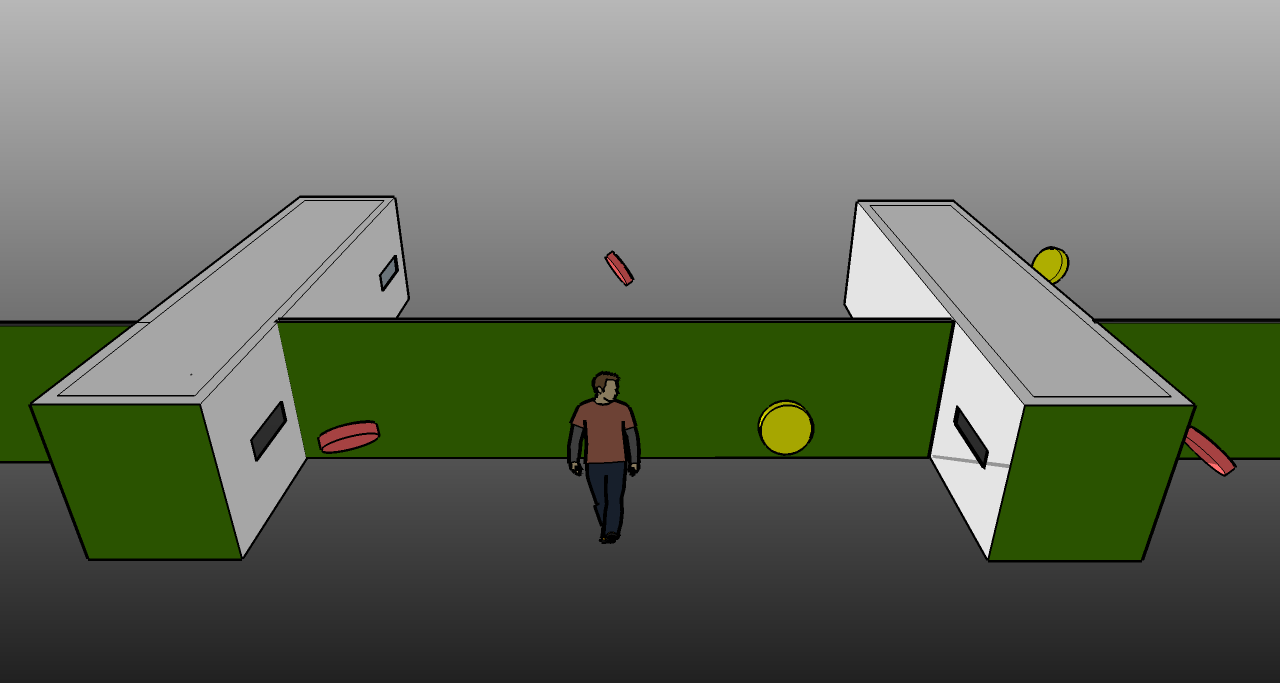}
\caption{\label{FIG1} \emph{Classical stochastic dynamics in presence of intermediate correlations.} Imagine a series of machines that take coins in as inputs and the output the coin. The above figure illustrates the middle of such a process. The experimenter can simply observer the value of the coin coming out of the left machine and let it be the input for the right machine. By observing the relative frequency of the outputs of the right machine conditioned on the outputs of left machine he can write down a stochastic map $\Lambda [\pd] = \qd$. However, as the back of the Figure shows the real system is made of two coins, whose dynamics is governed by a stochastic map $\Gamma [\pt] = \qt$. The observer in the front is oblivious to this fact as he cannot see the other coin. In fact, he simply see the the marginal distributions $\pd$ and $\qd$, founded by summing over the inaccessible degrees of freedom of $\pt$ and $\qt$ respectively. Now, note that the probabilities observed by the experimenter of the output of the right machine will depend on the value of the hidden coin on the left and therefore the stochastic map $\Lambda$ does not adequately describe the dynamics of the coin. One way to see that is that there is no easy way to obtain $\Lambda$ from $\Gamma$ alone. This essential difficulty can be entirely removed if our experimenter begins to alter the value of the coins coming out of the left machine. After observing the coin be output with probabilities $\pd$, the experimenter has the choice of flipping those coins so that they enter the second box with new probabilities $\pd'$. The combined action of the experimenter of the reading the output of the left machine and changing it to a different value is described by the action of a stochastic map $\xi$ acting on $\xi[\pd]=\pd'$. This is followed by the second part of the machine influencing these probabilities and the corresponding outputs being observed with a probability distribution $\qd$. In other words, the experimenter only has access the $\xi$ and observes the corresponding output $\qd$. Using these two he can only construct a map that transforms his choice $\xi$ to the probability distribution $\qd$. In the text, we demonstrate how the quantum version of this construction can be used to derive a second law that bounds the entropy production during the transformation that takes the choice of the experimenter to the next state. Given the correspondence between classical stochastic maps and completely positive quantum maps given in \cite{sudarshan1961stochastic}, this solution applies to the quantum mechanics just as well \cite{vinjanampathy2014second}.}
\end{center}
\end{figure}

Imagine that the system of coins, whose joint state is given by the probability distribution $\pt$ (see Fig. \ref{FIG1} for an illustration). The state of the first coin, $\pd$, is obtained as the marginal distribution from joint state of the two coins. We think of the first as the system, to which we have complete access, whereas the second coin represents inaccessible degrees of freedom that we will refer to as the \emph{environment}. Furthermore, imagine that the machine described before is in fact part of a bigger machine that accepts these two coins and outputs two coins. The total dynamics that the system is part of is described by a stochastic map $\Gamma$ acting on the two coin state: $\Gamma [ \pt ]= \qt $. We will observe the system in state $\pd = \tr_E [\pt]$ before the system coin entered the machine and $\qd = \tr_E [\qt]$ is the system state after leaving the machine. Here, $k$ represents the environmental index, and summing over it is the same as tracing over the second coin.

This dynamics, when viewed purely from the point of view of the system, is a transformation $\pd \rightarrow \qd$. Hence it is tempting to assign a stochastic map $\Lambda$ acting on the system, like before, such that $\Lambda[\pd]=\qd$. We can do this by observing the state of the system entering the machine and the corresponding output of the machine. The resultant map is derived from inverting Eq.~\eqref{Cstoch}. On the other hand we have 
\begin{gather}\label{fakemap}
\qd = \Gamma_{SE}[\pt_{SE}] = \left(\tr_E[\Gamma \, \pt_{E|S}] \right) [\pd]
\end{gather}
In the first equation the sum over $l$ is the trace over $E$ and in the second equation we have used conditional probability distribution $\pt = \pt_{E|S} \, \pd$. Therefore we have 
\begin{gather}
\Lambda = \tr_E[\Gamma \,  \pt_{S|E}]
\end{gather}
which includes the state of the environment conditioned on state of the system. Hence, the correlations present in $\pt$ prevent us from assigning such a map $\Lambda$ independently of the initial state of the system. In other words, when the system is observed in state `0' versus `1' the corresponding maps are different
\begin{gather}
\tr_E[\Gamma \, \pt_{E|0}] \ne \tr_E[\Gamma \, \pt_{k|1}].
\end{gather}
This is seen by noting $\pt_{E|S} = \td$ if and only if $\pt = \pd \otimes \td$, then
\begin{gather}
\Lambda = \Gamma \, \td.
\end{gather}
Above $\td = \sum_k t_k \ud_k$ is the local state of the second coin (environment). The presence of intermediate correlations, exemplified by the non-separability of the joint state $\pt$, causes the desired description of the system dynamics in terms of stochastic maps to fail. This discrepancy in describing dynamics with intermediate correlations in the quantum case is pointed out in \cite{modi2011preparation}. We will outline in the rest of this section, the solution to this problem of describing classical dynamics with intermediate correlations.

\subsection{Solution for classical processes}

The map constructed using Eq.~\eqref{fakemap} is problematic for other reasons too. It only predicts the output state $\qd$ corresponding to the observed input. But suppose we want to insert an input $\pd'$ of our choice. How can we construct a map that will predict the correct corresponding output?  The intermediate correlations present in the state of the classical system and the environment, namely $\pt$, disallowed the description of the state's dynamics in terms of stochastic maps. This issue is resolved if we take the view of an experimenter investigating the systems dynamics between the arrival of the coins and their entry into the machine. Such an experimenter has the ability to operate on the state of the system between the arrival of the coins and their subsequent entry to change it from $\pd$ to $\pd'$. We will demonstrate that this ability suffices to determine all aspects of the dynamics of interest to the experimenter.

Consider the stochastic matrices $\xi(\pd) = \pd'$, which does not act on the environment, i.e., 
\begin{gather}
\tr_\e[\xi \otimes \mathbf{I}(\pt)]= \tr_\e[\pt'] = \pd'. 
\end{gather}
We may think of $\xi$ to be a preparation procedure where $p'$ is the desired output. The corresponding output state of machine after such a preparation is given by
\begin{gather}\label{realmapind}
\qd = \tr_E[\Gamma [\xi \otimes \mathbf{I}  (\pt) ]]
= \tr_E[\Gamma \, \pt] [\xi] = \Theta [\xi]
\end{gather}
In other words the preparation map $\xi$ is the variable that allows us to alter the input to the machine. The mapping between $\xi$ and $q$ is given by $\Theta$:
\begin{gather}\label{realmap}
\qd = \Theta [\xi].
\end{gather}
We can construct $\Theta$ by choosing linearly independent preparation map $\xi^{(j,k)}$ and observing the corresponding output states $\qd^{(j,k)}$ (see below). It is important to keep in mind that $\Theta$ is not a mapping from $\pd'$ to $\qd$, rather from $\xi$ to $\qd$. This is because the mapping from $\pd \to \pd'$ is not give by an unique $\xi$. Eqs.~\eqref{realmapind} and \eqref{realmap} completely general and apply to arbitrary stochastic dynamics with small modifications. However, $\Theta$ is not a typical stochastic map. Rather it acts on stochastic maps and outputs a state. We will show how a second law can be derived for such processes.

Let us denote $\xi^{(j,k)}$ as a special set of operations that map the unit vector $\ud_j$ to $\ud_k$: 
\begin{gather}
\xi^{(j,k)}[\ud_l] = \ud_j (\ud_k \cdot \ud_l) = \ud_j \delta_{kl}
\end{gather}
The operations $\xi^{(j,k)}$ can be thought of as a measuring the system in state $\ud_k$ and the preparing it to $\ud_j$. We note that the dot product in the equation above is the usual vector dot product. The symmetric version $\xi^{(j,k)} + \xi^{(k,j)}$ is just a swap of the populations of $\ud_j$ and $\ud_k$. Now note that any transformation $\xi$ can be linear expanded as
\begin{gather}\label{linexp}
\xi = \sum_{jk} x^{(j,k)} \xi^{(j,k)},
\end{gather}
where $x^{(j,k)}$ are real expansion coefficients. This is possible since the operations $\xi^{(j,k)}$ form a linear basis on the space of stochastic maps acting on the system. Let us define action of $\Theta$ in Eq. \eqref{realmap} on $\xi^{(j,k)}$ is
\begin{gather}\label{baseop}
\Theta[\xi^{(j,k)}] = \qd^{(j,k)} \ud_j \delta_{kl}.
\end{gather}
Then using Eqs. \eqref{baseop} and \eqref{linexp} we can rewrite Eq. \eqref{realmap} as
\begin{gather}\label{realmap2}
\qd = \sum_{jk} x^{(j,k)} \qd^{(j,k)}.
\end{gather}
That is if we measure the finite set of distributions $\qd^{(j,k)}$ corresponding to the preparations $\xi^{(j,k)}$ 
then we can predict the output $\qd$ corresponding to any operation $\xi$.

\section{Second law}

To present a second-law like bound on the entropy production for stochastic processes with intermediate correlations, we begin by recollecting some basic properties of Stochastic maps. $d$-dimensional stochastic maps, like the preparation procedures considered above $\xi$, have the property that their entries are positive-semidefinite and each of their rows sum to 1. This allows us to arrange the rows of $\xi$ into a normalized vector $\xi^\uparrow$ producing a $d^2$-dimensional probability vector. Furthermore, with the action of every map $\Theta$, we can associate a map $\Theta^{\sharp}$, such that $\Theta^{\sharp} [\mathbf{\xi^\uparrow}]= \qd^\uparrow \equiv \qd \otimes \mathrm{id}/d$. Here $\mathrm{id}$ is an $d$-dimensional probability vector with entries $1$, and we note that $\Theta^{\sharp}$ now is a stochastic map that transforms $d^2$-dimensional probability vectors to $d^2$-dimensional probability vectors. To use this construction to describe the second-law like entropy bound, we have to appeal to the Kullback-Liebler divergence. The Kullback-Liebler divergence \cite{kullback1951information} of two probability distributions is defined as 
\begin{gather}
K(\xi^\uparrow \Vert \qd^\uparrow ) = - \xi^\uparrow \cdot(\log(\xi^\uparrow)-\log(\qd^\uparrow)).
\end{gather}

We can see that for diagonal density matrices, the quantum relative entropy $S(\rho\Vert\sigma)$ becomes the divergence measure defined above. Since stochastic maps are a subset of CPTP maps, we can immediately conclude that Kullback-Liebler divergence are contractive under stochastic maps. We can hence write $K(\mathbf{x} \Vert \mathrm{\mathbf{\epsilon}})\geq K( \Theta^{\sharp} [\xi^\uparrow] \Vert \Theta^{\sharp} [\mathrm{\mathbf{\epsilon}}])$. Here $\epsilon$ is a $d^2$ dimensional probability vector and the fixed point of the map $\Theta^{\sharp}$ (such a fixed point is guaranteed by Brower's theorem \cite{munkres1975topology}). Hence, owing to $\Theta^{\sharp}[\epsilon]=\epsilon$, we can write
\begin{align}\label{final}
H(\qd^\uparrow)-H(\xi^\uparrow) \geq - (\qd^\uparrow - \xi^\uparrow) \cdot \log{(\epsilon)},
\end{align}
where $H(\xi^\uparrow)$ is the Shannon entropy associated with the probability vector $\xi^\uparrow$. This is the second law for stochastic processes with intermediate correlations The law asserts that no transformations that violate Eq.~\eqref{final} are physically consistent with stochastic maps.

\section{Conclusions}
In conclusion, we have presented the correct entropy production bound for classical stochastic processes. The bound relied on the contractivity of Kullback-Liebler divergence under stochastic maps, which followed from the contractivity of quantum relative entropy under completely positive, trace preserving maps. We resolved the problem of describing the dynamics of systems whose probability vectors were correlated with an environment, which in turn was orchestrating the stochastic dynamics. The solution involved describing the dynamics in terms of $\Gamma$ maps, which describe the transformation of our choice of preparation of the input probability vector of the system to the output probability vector. Such a transformation relies on correctly describing the effect of conditioning the environmental dynamics, subject to measurements made on the system. We refer to \cite{vinjanampathy2014second} for the complete quantum generalization of this problem, and the corresponding generalized second law.

\section*{Acknowledgments}
Centre for Quantum Technologies is a Research Centre of Excellence funded by the Ministry of Education and the National Research Foundation of Singapore. We thank the organizers of the International Program on Quantum Information, 2013 held at the Institute of Physics--Bhubaneshwar, for the kind invitation to present a part of this work at the meeting.

\bibliographystyle{unsrt}
\bibliography{HSI-IJQI.bib}
\end{document}